\documentstyle{article}

\def\bichpr{\hoffset=-30truemm
\voffset=-25truemm
\textwidth=18 truecm
\textheight=24 truecm
}

\bichpr
\begin{document}
\title{Bethe-Salpeter equation: 3D reductions, heavy mass limits \\ and abnormal solutions. } 
\author{ J. Bijtebier\thanks{Senior Research Associate at the
  National Fund for Scientific Research (Belgium).}\\
 Theoretische Natuurkunde, Vrije Universiteit Brussel,\\
 Pleinlaan 2, B1050 Brussel, Belgium.\\
Email: jbijtebi@vub.ac.be}
%\date{}
\maketitle
\begin{abstract} \noindent  We show that the 3D reductions of the Bethe-Salpeter equation have the same bound state spectrum
as the original equation, with the possible exception of some solutions for which the corresponding 3D wave function vanishes. The
abnormal solutions of the Bethe-Salpeter equation (corresponding to excitations in the relative time-energy degree of freedom), when
they exist, are recovered in the 3D reductions via a complicated dependence of the
final potential on the total energy.  We know however  that the one-body (or one
high mass) limit of some 3D reductions of the exact Bethe-Salpeter equation leads to
a compact 3D equation (by a mutual cancellation of the ladder and crossed graph
contributions), which does not exhibit this kind of dependence on the total energy
anymore. We conclude that the exact Bethe-Salpeter equation has no abnormal
solution at this limit, or has only solutions for which our 3D wave function vanishes.
This is in contrast with the results of the ladder approximation, where no such
cancellation occurs. We draw the same conclusions for the static model, which
we obtain by letting the mass of the lighter particle go also to infinity.  These results
support Wick's conjecture that the abnormal solutions are a spurious consequence of
the ladder approximation. 
\end{abstract}
equations. \newline \noindent PACS 11.10.St \quad Bound and unstable states;
Bethe-Salpeter equations. \newline \noindent PACS 12.20.Ds \quad Specific calculations
and limits of quantum electrodynamics.\par  Relativistic bound states. Relativistic wave
equations. \\\\
%Short title: reblabla

\section{Introduction.}
  A well known difficulty of the Bethe-Salpeter equation \cite{1,2}  is the existence (for some interactions at
least) of abnormal solutions,  corresponding to excitations in the relative time-energy degree of freedom \cite{3,4,5,6}. Wick
suggested that these solutions could be an artefact due to  the ladder approximation, and would disappear if the higher-order irreducible
graphs could be included
\cite{3}. Mugibayashi showed  however that the abnormal solutions remain present in the exact static model \cite{7}, which is simple
enough to be completely solved.  Recently, the conclusions of Mugibayashi were questioned by Jallouli and Sazdjian \cite{8}, so that
the road to Wick's interpretation is
open again. This possible conspiration of the higher-order contributions for suppressing the abnormal solutions reminds us a 
similar cancellation of the higher-order terms at the one-body (or one large mass) limit of some 3D reductions of the Bethe-Salpeter
equation, leaving only a simple Dirac or Klein-Gordon equation \cite{9,10,11,12,13,14}.  A way of relating these two facts can be
found in our previous study of a zero-D reduction of the 1D Bethe-Salpeter equation for the static model at the ladder approximation
\cite{15}. We showed that the  reduction of the Bethe-Salpeter equation  does not in general kill the  possible abnormal solutions.
Despite the elimination of the relative time, they are recovered via the multiplicity of solutions  due to the complicated dependence of
the final  potential on the total energy.  This was the starting point of the present work.  It can be easily verified that this dependence of
the  reduced equation potential on the total energy is a necessary condition to the presence of abnormal solutions of any
reduction of a Bethe-Salpeter equation.  As this dependence disappears from some well chosen 3D
 reductions of the exact Bethe-Salpeter equation  at the one-body limit (and also from the zero-D reduction of the exact static model
obtained by letting the mass of the lighter particle go also to infinity), the 3D  equations have no abnormal solution at this limit. 
Conversely, these abnormal solutions are present in the reduction of the static model Bethe-Salpeter equation at the ladder
approximation, and we do not see why they would not be a fortiori present in the case of finite masses. \par
Does the absence of abnormal solutions for a 3D reduction imply their absence for the original Bethe-Salpeter equation itself? We
shall show that the existence of a solution of the Bethe-Salpeter equation for a given total energy implies in general the existence of
a solution of the reduced equation for this same total energy (there could be convergence problems, in principle manageable by
analytic extensions). However, the reduction being based on a
constraint fixing the value of the  relative energy,  a  solution of the Bethe-Salpeter equation will not lead to a corresponding  solution
of the reduced equation if it happens to vanish for this value. This accident does clearly not happen with the normal solutions of the
exact Bethe-Salpeter equation, and, from the study of the static model, we can deduce that it does not happen with at least some of the
abnormal solutions of the ladder approximated Bethe-Salpeter equation. For the possible abnormal solutions of the exact
Bethe-Salpeter equation, we have of course no example to examine (as we are just trying to demonstrate that such solutions do not
exist!). We can only conclude that these solutions, if they exist and survive at the one-body limit, are not transmitted to the 3D
reductions we use.\par In section 2, we present our 3D reduction of the exact or ladder approximated Bethe-Salpeter equation for two
fermions, the existence condition for this reduction and how to invert it. In section 3, we compute the one-body limit of this reduction in
the QED case, the one-body limit of the Bethe-Salpeter equation itself and verify that the reduction of the limit is the limit of the
reduction. In section 4, we get the static model by letting the mass of the former light fermion become also infinite, and compare again
the zero-D reduction of the zero-body limit with the limit of the reduction. We also verify that the existence condition for the reduction
is satisfied at least by the solutions of the ladder approximated Bethe-Salpeter equation for the static model. In section~5,  we consider
the Bethe-Salpeter equation and its 3D reduction for equal masses, and get the static model directly by letting this common mass go
to infinity. This leads to the same Bethe-Salpeter equation as in the preceding section, but to another zero-D reduction, based on a
different choice of the fixed relative energy.  Section 6 is devoted to conclusions.  
\section{The Bethe-Salpeter equation for two fermions and its 3D reduction.}
\noindent We shall write the Bethe-Salpeter equation  for the bound states
 of two fermions as:
\begin{equation}\Phi = G_0 K \Phi, \label{2.1}\end{equation} 
where $\Phi$ is the Bethe-Salpeter amplitude, function of the positions $x_1,x_2$ or of the momenta 
$p_1,p_2$ of the fermions, according to the representation chosen. The operator $K$ 
is the Bethe-Salpeter kernel, given in a non-local momentum representation by the sum of the irreducible
two-fermion Feynman graphs. The operator $G_0$ is the free propagator, given by the product
$G_{01}G_{02}$ of the two individual fermion propagators:
\begin{equation} G_{0i} = {1 \over p_{i0}-h_i+i\epsilon h_i}\,\beta_i = {p_{i0}+h_i\over
p_i^2-m_i^2+i\epsilon}\,\beta_i\label{2.2}\end{equation}
where the $h_i$ are the Dirac free hamiltonians
\begin{equation} h_i = \vec \alpha_i\, . \vec p_i + \beta_i\, m_i\qquad (i=1,2). \label{2.3}\end{equation}
The Bethe-Salpeter   kernel $\,K \,$ should contain charge renormalization and vacuum polarization graphs, while the
propagators $\, G_{0i}\,$ should contain self-energy terms (which can be transferred to $\, K\,$  \cite{16,17}). In this work, we
consider only the free fermion propagators in   $\,G_{0i}\,$ and the "skeleton" graphs in $\,K$, and we hope that the inclusion
of the various corrections would not change our conclusions.    We shall define the total (or external, CM, global)
and relative (or internal) variables:
\begin{equation} X = {1 \over 2} (x_1 + x_2)\ , \qquad P = p_1 + p_2\ , \label{2.4}\end{equation}
\begin{equation} x = x_1 - x_2\ , \qquad p = {1 \over 2} (p_1 - p_2),\label{2.5}\end{equation}
and give a name to the corresponding combinations of the free hamiltonians:
\begin{equation} S = h_1 + h_2\ , \quad s = {1 \over 2} (h_1 - h_2). \label{2.6}\end{equation}
In order to keep our formula's compact, we do not write the arguments of the functions and operators. When needed, we work in the
momentum representation (most often, we indicate only the relative energy explicitly).\par
\subsection{3D reduction with Breit's propagator.} 
\noindent The free propagator $G_0$ will be
approached by a carefully chosen expression
$G_\delta$, combining a constraint  fixing the relative energy, and a
global 3D propagator. The argument of the constraint
and the inverse of the propagator should be combinations of the operators used in the free
equations (at last approximately and for the positive-energy solutions).  We shall choose Breit's propagator \cite{18} 
\begin{equation}G_\delta=-2i\pi\, {1\over P_0-S}\, \delta(p_0-\mu)\beta_1\beta_2\label{2.7}\end{equation}
with
\begin{equation}\mu ={1\over 2P_0}(E_1^2-E_2^2) ={sS\over P_0}\label{2.8}\end{equation}
\begin{equation}E_i=\sqrt{h_i^2}=(\vec p_i^2+m_i^2)^{1\over 2}.\label{2.9}\end{equation}
We shall write the free propagator as the 3D approached propagator, plus a rest:
\begin{equation} G_0=  G_{\delta}+G_R.\label{2.25}\end{equation}
The Bethe-Salpeter equation  becomes then the inhomogeneous equation
\begin{equation}\Phi=G_0K\Phi=(G_\delta +G_R)K\Phi=\Psi +G_RK\Phi,\label{2.26}\end{equation}
with
\begin{equation}\Psi=G_\delta K\Phi \qquad (=G_\delta G_0^{-1}\Phi).\label{2.27}\end{equation}
Solving (formally) the inhomogeneous equation (\ref{2.26})  and putting the result into (\ref{2.27}), we
get
\begin{equation}\Psi=G_\delta K(1-G_RK)^{-1}\Psi=G_\delta K_T\Psi \label{2.28}\end{equation}
where
\begin{equation}K_T=K(1-G_RK)^{-1}=K+KG_RK+...=(1-KG_R)^{-1}K\label{2.29}\end{equation}
obeys
\begin{equation}K_T=K+KG_RK_T=K+K_TG_RK.\label{2.30}\end{equation}
The reduction series (\ref{2.29})  re-introduces in fact the reducible graphs into the Bethe-Salpeter   kernel,
but with $G_0$ replaced by $G_R$.
Equation (\ref{2.28})  is a 3D equivalent of the Bethe-Salpeter equation. It depends on the choice of
$G_\delta$ and is not manifestly covariant (of course, we can always write it in the center
of mass frame (assuming $P^2>0$) and then replace everything by its covariant equivalent).
We can write it as a pair of coupled equations by multiplying it by $P_0-S$ and $p_0-\mu$:
\begin{equation}(P_0-S)\Psi=-2i\pi \delta(p_0-\mu)\beta_1\beta_2 K_T\Psi,\label{2.31}\end{equation}
\begin{equation}(p_0-\mu)\Psi=0.\label{2.32}\end{equation}
The relative time can be completely eliminated by making
\begin{equation}\Psi\,=\,\sqrt{2\pi}\,\delta(p_0\!-\!\mu)\,\Psi'\label{2.33}\end{equation}
which gives
\begin{equation}(P_0-S)\Psi'\,=\,V'\Psi'\label{2.34}\end{equation}
\begin{equation}V'\,=\,-2i\pi\, \beta_1\beta_2\, K_T(\mu,\mu). \label{2.35}\end{equation}
The $\,P_0\,$ spectrum of (\ref{2.34}) should be that of the original Bethe-Salpeter equation (\ref{2.1}), without the
possible states for which the expression used in (\ref{2.27}) to define the wave function $\, \Psi\,$ vanishes. The normal bound state
spectrum, connected to the nonrelativistic spectrum, will be obtained at first order by keeping only the first term $\, K,$ (or an
approximation of it) in $\, K_T,$ the remaining of $\, K_T\,$ being treated as a perturbation. If the Bethe-Salpeter equation has also an
abnormal spectrum, it should, at least implicitly, remain present also in the 3D reduction (\ref{2.34}), via the $\, P_0-$dependence of $\,
V'.$ Writing
\begin{equation}P_0\Psi'\,=\,\left[S+V'(P_0)\right]\Psi'\label{2.36}\end{equation}             
and diagonalizing the hamiltonian
\begin{equation}\left[S+V'(P_0)\right]\Psi'_i(P_0)\,=\,E_i(P_0)\Psi'_i(P_0)\label{2.37}\end{equation}
we get the energy spectrum as the set of the roots of the equations 
\begin{equation} P_0-E_i(P_0)\,=\,0.\label{2.37a}\end{equation} 
For each value of $\, i,$ we should find 
the energy of the normal solution, plus possibly a set of abnormal energies. This scheme is in general too complicated to be used
in actual numerical calculations. We know three exceptions:\par 
--- When we want only to compute the normal solutions by perturbations.\par 
--- In the static model for two fermions in QED at the ladder
approximation (one photon exchange)
\cite{15}. In this case,
$\, (S+V')\,$ is not an operator anymore, it is simply a function of the energy and the coupling constant. The reduced equation (a
simple numerical equation) gives the only normal solution and the even abnormal solutions (for the odd abnormal solutions, $\,\Psi \,$
vanishes, as the relative energy is fixed to zero in the reduction performed in \cite{15} ).  The unique function
$\,E_i(P_0) \,$ is divided into sectors separated by poles and containing each one solution. The reduction series converges only in
the first sector containing the normal solution, but can be analytically extended by using the Pad\'e approximants method. \par
--- At the one-body or zero-body
limits of the exact equation, since the final 3D potential reduces then to a simple compact
term, as we shall see in section 3.
\subsection{Existence conditions and inversion of the reduction.} The 3D reduction performed above becomes impossible when the 3D
wave function $\, \Psi,$ defined by (\ref{2.27}), is the null function. This gives the
condition
\begin{equation}\Psi\,=\,G_\delta K \Phi\,\neq\,0\,\,\to\,\,(K\Phi)(\mu)\,\neq\,0\label{2.37b}\end{equation}
or, using the Bethe-Salpeter equation itself:
\begin{equation}\Psi'\,=\,{-2i\pi\over P_0-S}\,  \beta_1\beta_2 \,G_0^{-1}(\mu)\,\Phi(\mu)\,=\,
{-i\sqrt{2\pi}\over 4\,P_0^2}\,(P_0-S)(P_0^2-4\,s^2)\,\Phi(\mu)\,\neq\,0.  \label{2.38}\end{equation}
The expression
\begin{equation}(P_0-S)(P_0^2-4\,s^2)\,=\,(P_0-h_1-h_2)(P_0-h_1+h_2)(P_0+h_1-h_2),\label{2.39}\end{equation}             
when projected on the subspaces $\, h_i\!=\pm E_i,$ becomes a product of factors which can not vanish in the usual range of the
bound states $\,\vert m_1\!-\!m_2\vert<P_0<(m_1\!+\!m_2) .\,$ The reduction condition becomes thus simply $\, \Phi(\mu)\neq
0.\,$ The solutions  $\, \Phi(p_0)\,$ of the Bethe-Salpeter equation for which $\, \Phi(\mu)\!=\!0\,$ will thus not lead to solutions
of the 3D equation. This would be the case for the solutions antisymmetrical in $\,p_0 ,$ which may appear when the masses are
equal. We have of course the possibility of performing a reduction based on a different constraint, like putting a fermion on the
mass shell, as Gross \cite{9}.  The Bethe-Salpeter amplitude is given in terms of the wave function by
\begin{equation}\Phi(p_0)\,=\,\sqrt{2\pi}\,G_0(p_0)\,K_T(p_0,\mu)\,\Psi'.\label{2.40}\end{equation}
At $\,p_0\!=\!\mu ,$ we can factorize an operator $\, (P_0\!-\!S)^{-1}\,$ out of  $\,G_0(\mu) ,$ the 3D potential $\, V'\,$ out of $\,
K_T(\mu,\mu),$   and use the 3D equation (\ref{2.34})  to recover the relation (\ref{2.38}). 
\subsection{The QED case.} To be definite, we shall work in the QED framework, but our results will be more or less directly
adaptable to other interactions, and/or to the two-boson and boson-fermion systems \cite{13}. The modified kernel $\, K_T\,$
is then, in Feynman's gauge:
$$ K_T(p',p)={i\alpha\over4\pi^3}{1\over k^2+i\epsilon}(\gamma_1\! \cdot\!
\gamma_2) +\left({i\alpha\over4\pi^3}\right)^2\int \! dk_1\,{1\over k_2^2+i\epsilon}\,{1\over k_1^2+i\epsilon}\bigg [
(\gamma_1\! \cdot\! \gamma_2)\bigg(G_{01}(q_1)G_{02}(q'_{1B})$$
\begin{equation} +\big[{2i\pi\delta\over
P_0-S+i\epsilon}\big]_B\,\beta_1\beta_2\bigg)(\gamma_1\! \cdot\! \gamma_2)+\gamma_1^\mu
G_{01}(q_1)(\gamma_1\! \cdot\! \gamma_2)G_{02}(q'_{1C})\gamma_{2\mu}\bigg]+....\label{2.41}\end{equation}
The various contributions are symbolized by the graphs of figure 1.  In the reducible graphs,  we use $\, G_R\!=\!G_0\!-\!G_\delta\,$
instead of the $\,G_0 \,$ we would use in the calculation of the scattering amplitude, so that the graph B, for
example, symbolizes the sum of the second and third terms of (\ref{2.41}).  The external lines are there only to indicate the initial
momenta
$\, p_1,p_2\,$ and the final momenta $\, p'_1,p'_2\,$ of the fermions 1 and 2. The internal momenta of fermion 1 are $\,
q_1,q_2,...\,$ and the momenta of the photons leaving the fermion 1 line are $\, k_1,k_2,....$ All momenta will be written in
terms of the fixed external momenta and of $\,k_1,...k_{n-1},$ which we choose as integration variables. With these
conventions,      the different graphs of a given order differ only by the insertion points of the photon lines on the 
fermion 2 line. The internal momenta of fermion 2 will vary accordingly, and we shall denote them by $\,q'_1,q'_2,... ,$ with a
supplementary  index identifying the corresponding graph. The momenta of the intermediate fermion lines are thus:
\begin{equation} q_1=p_1-k_1, \qquad q_2=p_1-k_1-k_2, \qquad ...\qquad\hbox{all
graphs}.\label{2.42}\end{equation}
\begin{equation}q'_1=p_2+k_1, \qquad q'_2=p_2+k_1+k_2, \qquad ...\qquad\hbox{ladder graphs}.\label{2.43}\end{equation}
\begin{equation}n!-1 \,\,\,\hbox{ permutations of }\,\,k_1,...k_n \,\,\hbox{ in (\ref{2.43})  }\qquad\hbox{
crossed graphs}.\label{2.44}\end{equation}
\section{One-body limit.}
\subsection{One-body limit of the 3D reduction.} We shall now take the one-body limit $\, (m_2\!\to\!\infty)\,$ of the 3D
equation in QED.  We
shall first define E as a finite part of $P_0$:
\begin{equation} P_0=m_2+E.\label{3.1}\end{equation}
For $\mu$, we have, in any part of (\ref{2.41}):
\begin{equation} \mu\to-{m_2^2\over 2P_0}\approx{E-m_2\over 2}.\label{3.2}\end{equation}
This is also the asymptotic form to give to $p'_0$ and $p_0$ in (\ref{2.41}) , because of the constraint.
Combining this result with (\ref{3.1}), we get
\begin{equation} p'_{10},p_{10}\to E,\qquad p'_{20},p_{20}\to m_2.\label{3.3}\end{equation}
For the intermediate energies, we shall define more practical variables.
Considering, to be definite, the $\alpha^4$ diagrams, we have
\begin{equation}q_{10}=E-\omega_1, \qquad q_{20}=E-\omega_2,....\label{3.6}\end{equation}
\begin{equation}q'_{10}=m_2+\omega'_1, \qquad q'_{20}=m_2+\omega'_2,....\label{3.7}\end{equation}
\begin{equation}k_{10}=\omega_1-0,\qquad k_{20}=\omega_2-\omega_1,\qquad k_{30}=\omega_3-\omega_2,\qquad
k_{40}=0-\omega_3,\qquad \label{3.8}\end{equation}
\begin{equation}k_{40}=-(k_{10}+k_{20}+k_{30}).\label{3.9}\end{equation}
In the ladder diagram (figure 1):
\begin{equation}\omega'_1=k_{10},\qquad\omega'_2=k_{10}+k_{20},\qquad
\omega'_3=k_{10}+k_{20}+k_{30}.\label{3.10}\end{equation}
\begin{equation}\omega'_1=\omega_1,\qquad \omega'_2=\omega_2,\qquad
\omega'_3=\omega_3.\qquad\label{3.11}\end{equation} For the 4!-1 other diagrams of order $\alpha^4$, we have simply to
permute
$(k_{10},k_{20},k_{30},k_{40})$ in (\ref{3.10}).
This leads to an equation with only the $\,\beta_2\!=\!1 \,$ components, where $\, h_2\,$ can be replaced by  $\, m_2\,$  \cite{13}:
\begin{equation}\psi\,=\,{1\over E-h_1+i\epsilon h_1}\,V'\,\psi\label{3.11b}\end{equation}
$$V'(\vec p',\vec p)\,=\, -2i\pi \left\{{i\alpha\over4\pi^3}{-1\over
\vec k^2} +\left({i\alpha\over4\pi^3}\right)^2\int \! dk_1\,{1\over
k_2^2+i\epsilon}\,{1\over k_1^2+i\epsilon}\,\bigg[{1\over \omega_1+i\epsilon}+{2i\pi\delta(\omega_1)+{1\over \omega'_{1C}+i\epsilon}}
\bigg]\,G_{01}(q_1)\beta_1\right.$$
$$\left. +\left({i\alpha\over4\pi^3}\right)^3\int \! dk_2dk_1\,{1\over
k_3^2+i\epsilon}\,{1\over
k_2^2+i\epsilon}\,{1\over k_1^2+i\epsilon}\bigg\{\bigg[{1\over \omega_2+i\epsilon}+2i\pi\delta(\omega_2)
\bigg]\bigg[{1\over \omega_1+i\epsilon}+2i\pi\delta(\omega_1)
\bigg]\right.$$   
\begin{equation}\left.+...\bigg\}\,G_{01}(q_2)\beta_1G_{01}(q_1)\beta_1+...\right\}.\label{3.11c}\end{equation}
with
\begin{equation}G_{01}(q_i)\beta_1\,=\,{1\over E-\omega_1-h_1(\vec q_i)+i\epsilon h_1(\vec q_i)}\label{3.12}\end{equation}
\begin{equation}\vec q_1\,=\,\vec p_1-\vec k_1,\qquad \vec q_2\,=\,\vec p_1-\vec k_1-\vec k_2, ...\label{3.13}\end{equation}
Since $\,\omega'_{1C}\!=\!-\omega_1,$ we see that the second-order contributions cancel mutually.  It can be shown that similar
cancellations occur at each order \cite{9,10,11,12,13}, so that we remain with the first term of $\, V'\,$ (a Coulomb potential) only. 
This result can also be obtained in other gauges. Equation (\ref{2.37a}) has then only
one root, corresponding to the normal solution. At the ladder aproximation,
however, we have only the terms in 
\begin{equation}{1\over \omega_i+i\epsilon}+2i\pi\delta(\omega_i)\,=\,{1\over
\omega_i-i\epsilon}\label{3.14}\end{equation}
with nothing to cancel them.   Equation (\ref{2.37a}) can then have several roots. We have actually computed them at the zero-body
limit ($\,m_1\!\to\!\infty,$ static model) \cite{15}, and we do not see how they could not be present for a finite $\, m_1.$    It must be
noted that, if we make the instantaneous approximation without making the ladder approximation, the cancellation does not occur
either: we remain  with a dependence of the potential on the total energy, and possibly again with several roots of
(\ref{2.37a}).  
\subsection{One-body limit of the Bethe-Salpeter equation and its 3D reduction.} In order to get the $\, m_2\!\to\!\infty\,$ limit
of the Bethe-Salpeter equation (\ref{2.1}), we shall write 
\begin{equation}P_0\,=\,m_2+E,\qquad p_0\,=\,\mu-\omega.\label{3.15}\end{equation}
Writing
$\,\Phi(\mu\!-\!\omega)=\phi(\omega),\,\,K(\mu\!-\!\omega',\mu\!-\!\omega)=\kappa(\omega',\omega)$
we get, at the $\, m_2\!\to\!\infty\,$ limit:
\begin{equation}\phi(\omega)\,=\,{1\over E-h_1-\omega+i\epsilon h_1}\, {1\over
\omega+i\epsilon}\,\beta_1\,(\kappa\,\phi)(\omega). 
\label{3.16}\end{equation}
The 3D reduction of (\ref{3.16}) is obtained by writing
\begin{equation}\phi(\omega)\,=\,\psi(\omega)+{1\over E-h_1-\omega+i\epsilon
h_1}\, {1\over
\omega-i\epsilon}\,\beta_1\,(\kappa\,\phi)(\omega),
\label{3.17}\end{equation}
\begin{equation}\psi(\omega)\,=\,\sqrt{2\pi}\,\delta(\omega)\psi',\qquad \psi'\,=\, {-i\sqrt{2\pi}\over E-h_1+i\epsilon
h_1}\,\beta_1\,(\kappa\phi)(0).\label{3.18}\end{equation}
Here, $\,G_R  \,$ is obtained from $\, G_0\,$ by a simple change of the sign of one $\,i\epsilon \,$ (as it is also the case in
Gross reduction \cite{9}). The consequences of this change are not infinitesimal, as the integral of $\, G_R\,$  with respect to
$\,\omega\,$ is zero for the positive values of $\, h_1.$  Putting the iterations of (\ref{3.17}) in (\ref{3.18}) gives
$$\psi'\,=\,{-2i\pi\over E-h_1+i\epsilon h_1}\,\bigg[\beta_1\,\kappa(0,0)+\int\!d\omega_1\,\beta_1\kappa(0,\omega_1){1\over(
-\omega_1+ E-h_1+i\epsilon h_1)(\omega_1-i\epsilon)}\,\beta_1\,\kappa(\omega_1,0)$$ 
\[+\int\!d\omega_2d\omega_1\,\beta_1\kappa(0,\omega_2){1\over( -\omega_2+
E-h_1+i\epsilon h_1)(\omega_2-i\epsilon)}\,\beta_1\,\kappa(\omega_2,\omega_1)\]                  
\begin{equation}{1\over( -\omega_1+
E-h_1+i\epsilon h_1)(\omega_1-i\epsilon)}\,\beta_1\,\kappa(\omega_1,0)+...\bigg]\psi'\label{3.19}\end{equation}
With the exact kernel, the higher-order terms of $\, \kappa\,$ (crossed diagrams) cancel the higher-order terms of the expansion
(\ref{3.19}) (we get (\ref{3.11b}) with the first term of (\ref{3.11c})).  At the
ladder approximation, the kernel is given by
\begin{equation}\beta_1\kappa(\omega_2,\vec q_2;\omega_1,\vec q_1)\,=\,
{i\alpha\over4\pi^3}\,
\,{1\over(\omega_2\!-\!\omega_1)^2-(\vec q_2\!-\!\vec q_1)^2+i\epsilon} 
\label{3.20}\end{equation}
Equations
(\ref{3.19}-\ref{3.20}) give the same results as  Equations
(\ref{3.11b}-\ref{3.11c}), both for the exact model and for the ladder approximation.\par 
In order to perform this 3D reduction, we must have $\,\psi'\!=-i\sqrt{2\pi}(\omega\phi)(0)\neq\!0,$ (using (\ref{3.16}) and (\ref{3.18})).
This implies that
$\,
\phi(\omega)\,$ must have a pole at origin (accordingly, (\ref{2.38}) shows that $\, \Phi(\mu)\,$ must be infinite at the
$\,m_2\!\to\!\infty \,$ limit if we want to get a nonvanishing $\, \Psi'\,$). This is certainly true for the normal solutions,
for which we can in principle compute $\psi'$ by perturbations around a nonrelativistic approximation. We shall verify below that it is
also true for the abnormal solutions  of the ladder approximated Bethe-Salpeter equation at the zero-body
$\,(m_1\!\to\!\infty) \,$ limit.
\section{Asymmetrical zero-body limit.}      
\subsection{Asymmetrical zero-body limit of the 3D reduction.}    
The $\,m_1\!\to\!\infty \,$ limit of the reduction (\ref{3.11b}-\ref{3.11c}) will be performed after writing $\, E\!=\!m_1\!+\!W$. 
At the ladder approximation, we get
$$W\psi'(\vec p'_1)\,=\,-2i\pi\int\!d^3\!p_1\bigg[ {i\alpha\over4\pi^3}\, {1\over-(\vec p'_1\!-\!\vec p_1)^2+i\epsilon}$$$$ 
+ \,\left({i\alpha\over4\pi^3}\right)^2\int \!d^3\!q_1d\omega_1\,{1\over\omega_1^2-(\vec p'_1\!-\!\vec
q_1)^2+i\epsilon}\,\,{1\over(-\omega_1+W+i\epsilon)(\omega_1-i\epsilon)}\,\,{1\over\omega_1^2-(\vec q_1\!-\!\vec
p_1)^2+i\epsilon}$$   
$$ 
+\, \left({i\alpha\over4\pi^3}\right)^3\int \!d^3\!q_2d\omega_2d^3\!q_1d\omega_1\,{1\over\omega_2^2-(\vec p'_1\!-\!\vec
q_2)^2+i\epsilon}\,\,{1\over(-\omega_2+W+i\epsilon)(\omega_2-i\epsilon)}\,\,$$   
\begin{equation}{1\over(\omega_2\!-\!\omega_1)^2-(\vec q_2\!-\!\vec
q_1)^2+i\epsilon}\,\,{1\over(-\omega_1+W+i\epsilon)(\omega_1-i\epsilon)}\,\,{1\over\omega_1^2-(\vec q_1\!-\!\vec
p_1)^2+i\epsilon}\,+...\bigg]\,\psi'(\vec p_1).\label{4.2}\end{equation}
With the exact kernel, we must keep only the first term. Since we expect to get the static model, we shall search for solutions in
\begin{equation}\psi'(\vec p_1)\,=\,\exp[\,i\vec p_1\cdot\vec x]\label{4.3}\end{equation}
where $\,\vec x \,$  is a set of three continuus quantum numbers which can be interpretated as the conserved
relative position of two infinitely heavy fermions. As  the
only remaining 3-momenta are $\, \vec k_1\!=\!(\vec p_1\!-\!\vec q_1),\,\vec
k_2\!=\!(\vec q_1\!-\!\vec q_2),\,$ etc.... in the   propagators of the photons we shall
perform the integrations with respect to the $\, \vec k_i\,$.  Using the relations
\begin{equation}\vec p_1\,=\,\vec p'_1-(\vec k_1+\vec k_2+...),\label{4.3b}\end{equation}
\begin{equation}\int\!d^3\!k \,\,{\exp[\,i\,\vec k\cdot\vec x]\over\omega^2-\vec k^2+i\epsilon}\,=\,
{-2\pi^2\over\vert\vec x\vert}\,\exp[\,i\vert\omega \vec x\vert],  \label{4.4}\end{equation}
   removing a common factor $\,\exp[\,i\vec p'_1\!\cdot\!\vec x]
\,$ and choosing  $\,\vert \vec x \vert\!=\!1,$ which means taking $\,1/\vert\vec x\vert\ \,$ as the
energy unit and rendering $\, W\,$ and $\,\omega \,$ dimensionless (the energy will
thus be $\,W/\vert\vec x\vert$ from now on), we get
$$W\,=\,-2i\pi\bigg[ {-i\alpha\over2\,\pi}\,+\, \left({-i\alpha\over2\,\pi}\right)^2\int\!d\omega_1\,\exp[i\vert
0\!-\!\omega_1\vert]\, {1\over(-\omega_1+W+i\epsilon)(\omega_1-i\epsilon) }\,\exp[i\vert
\omega_1\!-\!0\vert]$$$$+\, \left({-i\alpha\over2\,\pi}\right)^3\int\!d\omega_2d\omega_1\,\exp[i\vert
0\!-\!\omega_2\vert]\, {1\over(-\omega_2+W+i\epsilon)(\omega_2-i\epsilon) }\,\exp[i\vert
\omega_2\!-\!\omega_1\vert]\, $$     
\begin{equation}{1\over(-\omega_1+W+i\epsilon)(\omega_1-i\epsilon) }\,\exp[i\vert
\omega_1\!-\!0\vert]+...\bigg] \label{4.5}\end{equation} 
With the exact kernel, we have simply $\, W\!=\!-\alpha.$ \par    
\subsection{Zero-body limit of the Bethe-Salpeter equation and asymmetrical zero-D reduction.}    
The $\,m_1\!\to\!\infty \,$ limit of the Bethe-Salpeter equation  (\ref{3.16}) will be, after writing $\, E\!=\!m_1\!+\!W$:
\begin{equation}\phi(\omega',\vec p'_1)\,=\, {1\over(W-\omega'+i\epsilon)(\omega'+i\epsilon)}\, 
\int\!d^3\!p_1\,d\omega\,\, \kappa(\omega',\vec p'_1;\omega,\vec p_1)\,\phi(\omega,\vec p_1).  
\label{4.6}\end{equation}  We shall search for solutions in
\begin{equation}\phi(\omega,\vec p_1)\,=\,\chi(\omega)\,\exp[\,i\,\vec p_1\cdot\vec x].\label{4.7}\end{equation} 
The only 3-momenta in $\,\kappa\,$ being again the $k_i$\, of the photon propagators (in both the exact and the ladder approximated
models), we shall again perform the integrations with respect to the $\,k_i.$\, Defining 
\begin{equation}
\int\!d^3\!p_1\,\, \kappa(\omega',\vec p'_1;\omega,\vec p_1)\,\exp(i\vec p_1\cdot\vec x)\,=\,\exp(i\vec p'_1\cdot\vec x)\,
\eta (\omega',\omega).  
\label{4.7b}\end{equation} 
 and choosing $\,\vert \vec x \vert\!=\!1,$ we get  
\begin{equation}\chi(\omega')\,=\, {1\over(W-\omega'+i\epsilon)(\omega'+i\epsilon)}\, \,
\int\!d\omega\,\, \eta(\omega',\omega)\,\chi(\omega).   \label{4.8}\end{equation}
At the ladder approximation, we have
\begin{equation}\kappa(\omega',\vec p'_1;\omega,\vec p_1)\,=\,{i\alpha\over4\pi^3}
\,\,{1\over (\omega'-\omega)^2-(\vec p_1'-\vec p_1)^2+i\epsilon},\qquad  
\eta(\omega',\omega)\,=\,{-i\alpha\over 2\pi}\exp(i\vert\omega'-\omega\vert).\label{4.8b}\end{equation}
The zero-D reduction  is obtained, as in section 3.2, by changing the sign of the second $\, i\epsilon\,$ of (\ref{4.8}), leaving
a term
 in $\,\delta(\omega') \,$ as the initial term of an iteration. This gives (\ref{4.5}) at the ladder approximation, and\,
$W\!=\!-\alpha\,$ for the exact model. The condition for this reduction is      
\begin{equation}\int\!d\omega\, \,\eta(0,\omega)\,\chi(\omega)\,\neq\,0   \label{4.9}\end{equation}
which implies, via (\ref{4.8}), that $\, \chi(\omega)\,$ must have a pole at origin. 
\subsection{Condition for the asymmetrical zero-D reduction of the ladder-approximated static model.}
We think that the absence of abnormal solutions of the 3D reduction of the exact Bethe-Salpeter equation at the one-body limit is
probably the consequence of their absence for the original finite-masses Bethe-Salpeter equation itself. There remains however the
possibility that such solutions exist and disappear as a consequence of the one-body
limit, of the 3D reduction, or of the combination of both.  It is of course impossible
to study directly the behaviour of these solutions, the very existence of which is
questioned.  We know anyway that the abnormal solutions of the Wick-Cutkosky
model survive the one-body limit \cite{55}. We know also that the abnormal
solutions of the ladder-approximated static model survive the zero-D reduction we
performed in \cite{15}. This last zero-D reduction is however based on the choice
$\,\omega\!=\!W/2\,$ (just between the two poles at $\,\omega\!=\!0\,$ and
$\,\omega\!=\!W)\,$ and treats the two fermions in a symmetrical way, while the present zero-D reduction is based on the choice
$\,\omega\!=\!0\,$, in order to get the very  compact result $\,W\!=\!-\alpha \,$ in the
exact model. We must thus check if the condition for this new zero-D reduction is
also satisfied by the abnormal solutions of the ladder approximated static model. If we
write
\begin{equation}u(\omega)\,=\,\chi(\omega+ {W\over2}), \label{4.10}\end{equation}
we get a more symmetrical Bethe-Salpeter equation:
 \begin{equation}u(\omega')\,=\, {1\over  {1\over4} W^2-\omega'^2-i\epsilon}\, \,{-i\alpha\over2\pi}
\int\!d\omega\, \exp\,[i\vert\omega'\!-\!\omega\vert]\,u(\omega).   \label{4.11}\end{equation}
where the sign before the $\, i\epsilon\,$ comes from the fact that we are interested in negative values of $\, W\,$ (we had
$\,+i\epsilon W).\,$  Equations (\ref{4.8}) or  (\ref{4.11}) are the Bethe-Salpeter equation for the static model, at the ladder
approximation, in the relative energy representation. In a previous work \cite{15}, we studied  equation (\ref{4.11}) in the relative time
representation, performing a Wick rotation and a zero-D reduction. We could meet our previous results by performing a Fourier
transform of (\ref{4.11}), but it turns out to be also possible without leaving the  relative energy representation. Writing  
\begin{equation}u(\omega)\,=\, {1\over  {1\over4} W^2-\omega'^2-i\epsilon}\, \,v(\omega),  \label{4.12}\end{equation}
\begin{equation}W=-2w,\qquad \omega=-wt,\qquad {\alpha\over2\pi}=\lambda, \label{4.13}\end{equation}
we get 
\begin{equation}v(t')\,=\, {i\lambda\over w}
\int\!dt\, {\exp\,[iw\vert t'\!-\!t\vert]\over t^2-1+i\epsilon}\, \,\,v(t).  
\label{4.14}\end{equation}
The condition for the zero-D reduction (\ref{4.5}) is that $\, v(t)\,$ must not vanish at $\,t\!=\!-1.$ If we derive twice equation
(\ref{4.14}), we get the differential equation
\begin{equation}v''(t)\,+\, \left[{2\lambda\over t^2-1}\,+\,w^2\right]v(t)\,=\,0. \label{4.15}\end{equation}
Although we are still in the relative energy representation, equation (\ref{4.15}) and the complex conjugated equation (\ref{4.14})
(written thus for
$\, v^*\,$) are exactly the equations we studied previoulsly in the relative time representation \cite{15} (the Fourier transform of the
propagator takes the form of the potential, and vice-versa). We found that we had to solve the differential equation (\ref{4.15}), and,
using the Wick rotation in (\ref{4.14}), that the solutions had to be square integrable along the imaginary axis. Equation (\ref{4.15})
being regular and symmetric around the origin, the solutions are regular at origin and can be made even or odd. In fact, for each
eigenvalue $\, w_i,$ we found either an even or an odd solution, but not both simultaneously. A zero-D reduction of (\ref{4.11}),
with             
 \begin{equation}G_0(\omega)\,=\, {1\over  {1\over4}
W^2-\omega^2-i\epsilon},\qquad 
G_\delta(\omega)\,=\, \delta(\omega)\int{d\omega'\over  {1\over4} W^2-\omega'^2-i\epsilon}\,
=\,-2i\pi
\,{\delta(\omega)\over W}   \label{4.16}\end{equation} demands $\, v(0)\!\neq\!0,$ so that the odd solutions of the Bethe-Salpeter
equation were not solutions of the reduced equation (we did not miss them, as they do not satisfy the normalisation condition deduced,
like the homogeneous Bethe-Salpeter equation for the amplitudes, from the inhomogeneous Bethe-Salpeter equation for the
propagator). \par To see if  a solution of
(\ref{4.14})  leads to a solution of the reduced equation (\ref{4.5}), we must verify that it does not vanish at $\, t\!=\!-1.$ The
solutions $\, v\,$ of (\ref{4.15}), a Fuchs equation around this point, can be written as linear combinations of two solutions of the form
\begin{equation}v^+=\sum _{n=0}^\infty C^+_n(t+1)^{n+1}, \qquad v^-=\sum _{n=0}^\infty
C^-_n(t+1)^n\,+\,c\,v^+\,\log(t+1)\label{4.17}\end{equation}
in the  $\,\vert  t\!+\!1\vert<2$ domain of the $\, t\,$ complex plane. At  $\, t\!=\!-1,$ $\, v^+\,$ is zero, and $\, v^-\,$ is $\,
C^-_0,$ by definition a nonzero arbitrary constant. As written above, we know also that we must build even and odd  solutions around
$\, t\!=\!0.$  In order to build these solutions, we must in general use a linear combination of $\, v^+\,$ and  $\,v^- ,$ which will differ
from zero at
$\, t\!=\!-1,$ unless built with 
 $\,v^+ \,$ alone, which should then be  even or odd when re-expanded into a series of powers of $\,t.$  Although this would imply an
infinity of cancellations between the coefficients $\,C^+_n ,$ we did not exclude this possibility a priori. We checked numerically
that it does not happen for the 20 first eigenvalues computed in ref. \cite{15}.
\section{Symmetrical zero-body limit.}                 
\subsection{Symmetrical zero-body limit  of the 3D reduction.}
We can take directly the $\,m_1\!=\!m_2\!=\!m\!\to\!\infty \,$ limit of the 3D reduction. We have and keep $\, \mu\!=\!0.$
We shall write
\begin{equation}P_0\,=\,2m+W,\label{5.1}\end{equation} 
\begin{equation}p_{10}=p_{20}=p'_{10}=p'_{20}\,\to\,m+ {W\over2}, \label{5.2}\end{equation} 
\begin{equation}q_{10}\,=\,m+ {W\over2}-\omega_1,\qquad q_{20}\,=\,m+ {W\over2}-\omega_2, \label{5.3}\end{equation} 
\begin{equation}q'_{10}\,=\,m+ {W\over2}+\omega'_1,\qquad q'_{20}\,=\,m+
{W\over2}+\omega'_2.\label{5.4}\end{equation}  The fermion propagators, in the second-order terms of the exact model, will give a
contribution proportional to
\begin{equation}\left[ {1\over {1\over2}W+\omega_1+i\epsilon }\,+\,i\pi\delta(\omega_1)\,+\,{1\over
{1\over2}W+\omega'_{1C}+i\epsilon }
\right]\,{1\over {1\over2}W-\omega_1+i\epsilon }\label{5.5}\end{equation}
 with $\,\omega'_{1C}\!=\!-\omega_1 \,$ (at the ladder approximation, we would not get the crossed term). In contrast with what
happens with the asymmetrical zero-body limit, these terms do not combine to zero. In Coulomb's gauge, however,  the only
relative energy dependence lies in the fermion propagators, and the integral of (\ref{5.5}) with respect to $\, \omega_1\,$ is zero. The
integral of the crossed term is indeed zero, as the poles lie both on the same (upper) half part of the $\, \omega_1$-complex plane. The
integrals of the two other contributions combine to zero (they correspond to the modified ladder term, $\, G_R,$ which was designed
for that). At higher orders, we find the products of the lower fermion propagators $\, (W/2\!+\!\omega'_{iG}\!+\!i\epsilon)^{-1},$
where $\,
\omega'_{iG}\,$ corresponds to the ith lower fermion propagator of the graph $\,G.$ It can be shown that the $\,
\omega'_{iG}\,$ are combinations of the $\, \omega_i,$ with coefficients $\,\pm1,0 .$ For the uncorrected ladder graphs, $\,
\omega'_{iG}\!=\!\omega_i.$ For the crossed graphs, at least one $\, \omega_i\,$ appears only with minus signs in the
lower fermion propagators, so that the integral with respect to this $\, \omega_i\,$ will be zero. We remain then with the
modified ladder graphs (built with the ladder term plus the $\, \delta\,$) which give also zero by integration. We
conclude that, in Coulomb's gauge, both the exact kernel and the ladder approximation give the normal solution only.
This confirms our result above (and that of \cite{8}) about the absence of abnormal solutions in the exact static model,
without contradiction with our result about the ladder approximation, which is not supposed to be gauge-invariant.
\subsection{Zero-body limit of the Bethe-Salpeter equation and symmetrical zero-D reduction.}                  
Writing
\begin{equation}P_0=2m+W,\qquad p_0=-\omega\label{5.6}\end{equation} 
in the Bethe-Salpeter equation at
the ladder approximation, and taking the $\, m\!\to\!\infty\,$ limit, we get directly the symmetrical equation (\ref{4.11}). The
zero-D reduction of (\ref{4.11}), using (\ref{4.16}), gives    
$$W\,=\,-2i\pi\bigg[ {-i\alpha\over2\,\pi}\,+\, \left({-i\alpha\over2\,\pi}\right)^2\int\!d\omega_1\,\exp[i\vert
0\!-\!\omega_1\vert]\, \left\{{1\over  {1\over4} W^2-\omega_1^2-i\epsilon}+2i\pi \,{\delta(\omega_1)\over W}\right\}\,\exp[i\vert
\omega_1\!-\!0\vert]$$$$+\, \left({-i\alpha\over2\,\pi}\right)^3\int\!d\omega_2d\omega_1\,\exp[i\vert
0\!-\!\omega_2\vert]\, \left\{{1\over  {1\over4} W^2-\omega_2^2-i\epsilon}+2i\pi \,{\delta(\omega_2)\over W}\right\}\,\exp[i\vert
\omega_2\!-\!\omega_1\vert]\, $$\begin{equation}\left\{{1\over  {1\over4} W^2-\omega_1^2-i\epsilon}+2i\pi
\,{\delta(\omega_1)\over W}\right\}\,\exp[i\vert
\omega_1\!-\!0\vert]+...\bigg] \label{5.7}\end{equation} 
The terms between brackets correspond to the two first terms (modified ladder part) of (\ref{5.5}).
\section{Conclusions.}
We pointed out that the cancellation mechanism which transforms some well chosen 3D reductions of the exact Bethe-Salpeter
equation into a simple Dirac equation (for two fermions in QED) at the one-body limit, implies  that these reductions have no abnormal
solution at this limit, nor at the zero-body limit (static model). Conversely, this cancellation does not occur at the ladder approximation,
so that the abnormal solutions are present in this case. Our argumentation is based on the fact that the abnormal solutions, when present,
are implicitly preserved by the 3D reduction, via the dependence of the potential on the total energy, and should thus disappear when
this potential reduces to a simple compact potential, like a Coulomb potential. 
This cancellation mechanism in the 3D reductions at the one-body limit occurs in many situations (2 fermions, 2 bosons, 1 boson +
1 fermion with various interactions) \cite{9,10,11,12,13,14}, providing many new examples for the discussion of the abnormal
solutions problem. The 3D reduction must be based on a constraint which puts, at the one-body limit, the heavy particle on its mass
shell. Putting directly one particle on its mass shell, as in the 3D reduction of Gross \cite{9} without taking the one-body limit, is
however not enough. These facts  and the result of ref. \cite{8} about the absence of abnormal solutions in the exact static model,
support Wick's conjecture that the abnormal solutions are  spurious consequences of the ladder approximation,  and would
disappear in the exact model. However, it remains possible that the exact model has
abnormal solutions, which are killed by the one-body limit, the 3D reduction (if the
corresponding Bethe-Salpeter amplitudes vanish for the relative energy fixed in the
reduction), or the combination of both. This does clearly not happen with the normal
solutions, and our study of the static model has shown that it does not happen either
with the abnormal solutions at the ladder approximation. 
\par The main problem of the existence of abnormal solutions of the exact, finite masses Bethe-Salpeter equation remains thus open. 
 What we have shown in this work is 
that the higher-order contributions to the 3D potential of well chosen reduced equations for the exact model cancel mutually at leading
order in each of the masses. The dependence on the total energy, which is related to the presence of abnormal solutions, is thus much
less important than at the ladder approximation, and possibly too small to allow for more than one root of $\, P_0\!-\!E_i(P_0).$  It
could perhaps be possible to eliminate this dependence completely order by order, using the large freedom which we still have in the
choice of the 3D reduction.  We made clear anyway that, although  the non-ladder contributions can be treated as  correction terms in
the study of the normal solutions, they  modify completely (if not annihilate) the abnormal solutions obtained at the ladder
approximation.  The detection and computation of abnormal solutions goes thus beyond the validity of the ladder approximation.   
\par\vskip5mm\noindent {\bf Aknowledgements.} The author aknowledges the Belgian federal government's support (UIAP
46). \par  
\end{document}